\documentstyle[prl,aps,epsfig,multicol]{revtex}

\def\bfg{\begin{figure}}
\def\efg{\end{figure}}
\def\vr{\varrho}

\begin{document}
\title{Self-organization, Localization of 
Shear Bands and Aging in Loose Granular Materials} 

\author{J\'anos T\"or\"ok$^{1,2}$, Supriya Krishnamurthy$^2$,
J\'anos Kert\'esz$^1$ and St\'ephane Roux$^3$}
\address{$^1$Department of Theoretical Physics, 
Institute of Physics,\\ Technical University of Budapest, 
Budafoki \'ut 8, Budapest, H-1111, Hungary \\
$^2$Laboratoire de Physique et M\'ecanique des Milieux
H\'et\'erog\`enes,\\ 
ESPCI, 10 rue Vauquelin, Paris 75231, France.\\
$^3$Surface du Verre et Interfaces, UMR CNRS/Saint-Gobain, \\
39 Quai Lucien Lefranc, 93303 Aubervilliers Cedex, France}

\date{\today}
\maketitle

\begin{abstract}
We introduce a mesoscopic model for the formation and evolution
of shear bands in loose granular media. Numerical simulations reveal
that the system undergoes a non-trivial self-organization process
which is governed by the motion of the shear band and the consequent
restructuring of the material along it. High density regions are built
up, progressively confining the shear bands in localized regions. This
results in an inhomogeneous aging of the material with a very slow
increase in the mean density, displaying an unusual glassy like
system-size dependence.
\medskip

\end{abstract}
\begin{multicols}{2}
\narrowtext

A large class of materials are handled in the form of dispersed
solid grains at some stage of their processing. Thus the description
of the rheological properties of suspensions, pastes and dry granular
media is a key  question which controls the ability of 
mixing, storing, transporting etc, these disperse media
\cite{BidHan,PaG97,HovHer}. Granular systems constitute an
intermediate state of matter between fluids and solids
\cite{Wood,JNB}: they flow like fluids but they also build piles
indicating that a non-vanishing static shear stress is present which is
characteristic of solids.  From this point of view it is also of
major interest to understand the shearing process in these systems.
A number of experiments have been carried out on the shear process in
granular materials \cite{Rey,WoBu}. Most of these are triaxial tests
\cite{WoBu,Krenk} to determine macroscopic properties such as the
shear stress or the volumetric strain, as a function of the shear
strain.

The intimate interplay between the geometrical arrangements and the
frictional properties of the grains  determines the precise form of the
rheological behavior to be used  at a continuum level. The underlying
question is the identification of relevant internal variables. The most
obvious one is the density of the sample, which can be made to 
vary over a wide range by the method of preparation. 
Compared to other parameters
describing the texture (e.g. fabric tensors accounting for the
distribution of contact orientations) the density has the most drastic
impact on the stress needed to shear the material as well as on the mode
of shearing; from an apparent homogeneous strain for loose packings to a
localized steady shear band for dense assemblies \cite{DCMM}. The
coupling of the density to the shear properties can be understood
through the concept of dilatancy \cite{Rey}.

A related question is whether statistical fluctuations have an impact
on macroscopic properties. Lately, there has been an upsurge of
interest in trying to characterize the large stress fluctuations
\cite{Moldin,CBN,NasLac,Couette} in silos, Couette flow or slider
block geometries, or to understand the statistics of interparticle
contact forces \cite{LRR}. Recently, spectacular experiments in
two-dimensional Couette shear cells were carried out \cite{Couette}
where the movement and stress of individual particles were monitored
in order to describe the inner structure and the force network in the
sheared granular material. It was demonstrated that stationary motion
is accompanied by large stress fluctuations due to the formation and
breakdown of arches.  Large fluctuations were also found in three
dimensional steady state shear cells \cite{Behr}.

This issue has also been raised by the results of recent numerical
simulations of rigid grain assemblies \cite{More}, where even at low
densities, the shearing which appears as homogeneous over long times,
in fact consists of a succession of sudden changes of quasi-instantaneous
and localized strain fields. This observation suggests that the
transition from the particle based description to the continuum one
requires the detailed understanding of the statistical features
associated with these sudden changes.

In this Letter we present a simple model for the shearing of a
granular medium in loose samples. We describe the strain field at
every instant as a shear band, chosen through a global optimization
procedure, which is equivalent, as we shall see later, to searching
for the ground state of a 
directed polymer in a random potential \cite{dirpol,Maloy}. However,
this potential is not {\it a priori} frozen in but has a
self-organized development due to our procedure of choosing and
changing the shear band. Though very simple and with only the minimum
of ingredients, the model shows that the density of the
medium increases anomalously slowly. Further we are also able to
predict on the basis of this model, that 
large scale inhomogeneities build up in a system subject to a steady
shear. This could be an interesting feature to compare with experiments.

Let us consider a shear process, assumed to be invariant along the
shear direction  ($z$ in Fig. \ref{Fig_descr}).  This geometry is
appropriate for instance, in an annular shear cell of large radius 
\cite{Behr}.  We consider moreover a continuum
\bfg
\centerline{\epsfig{file=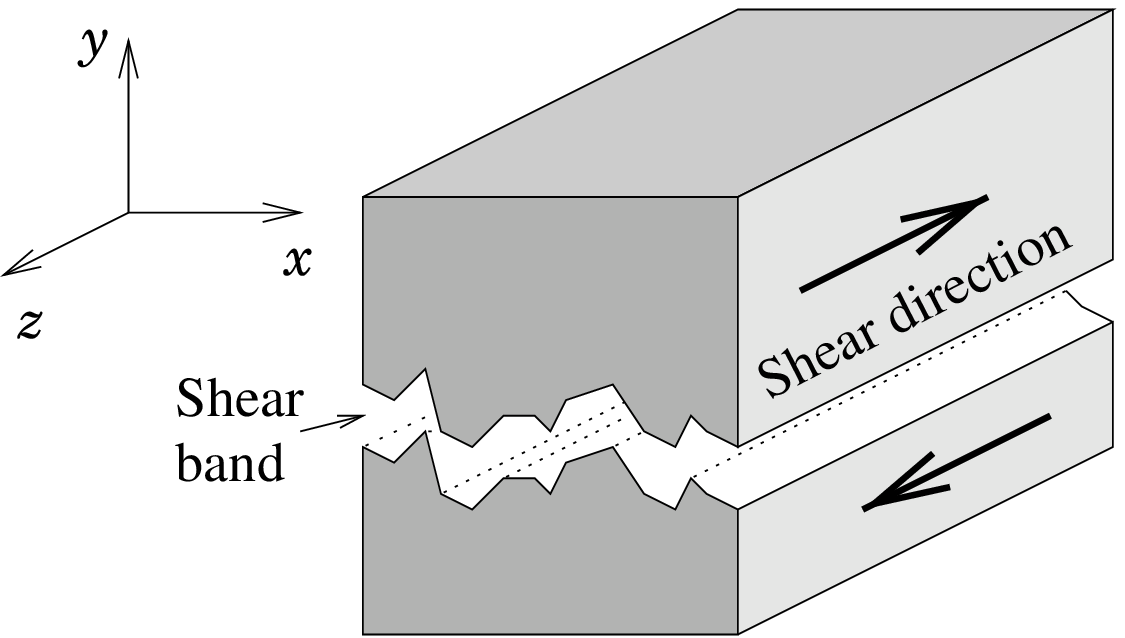,height=4truecm}}
\medskip
\caption{\label{Fig_descr}\footnotesize
Schematic picture of the shear process. The shear band is parallel to
the shear direction $z$ due to periodic boundary conditions in
this direction.
}
\efg\noindent\vskip-10pt
description, valid on scales much larger than
that of individual grains. We now introduce a
fundamental assumption of our model: We assume that the {\it
instantaneous} strain field is always localized on a single shear band
\cite{Wood,ToRo}. 
Experimentally, it is known that shear bands have a typical width of 
about ten grain diameters. Thus at a continuum level, the velocity
field is indeed discontinuous across the shear band. 
From the geometry of our set-up, the shear
band must be a continuous surface due to topological constraints
(Fig. \ref{Fig_descr}).  
Further, we assume that because of the translational
invariance along the $z$ axis, the system can  
be reduced to a 
two dimensional one in the $x$-$y$ (cross-section) plane, through an
averaging over the $z$ direction.  

The basic hypothesis of the localization of the shear on the
shear band at all times, is 
not as restrictive as it may appear. We only
refer here to instantaneous shear rates, and provided the shear band 
changes rapidly enough, coarse-graining  
the strain field in time may produce a uniform shear rate.   
Experimentally, though it is very
difficult to have direct access 
to the instantaneous shear rate, 
large fluctuations found in the shear stress may indicate that the shear is
never quite uniform, even at early times. As mentioned earlier,
this seems indicated also by numerics \cite{More}.

Initially we consider a loose-packed sample. At a suitably
coarse-grained scale the medium can be described as a continuum, where
the density is a random function displaying fluctuations around a mean
value. Under a constant normal load, a threshold shear force (or torque
for an annular shear cell) has to be applied to impose a non-zero strain.
Locally, after integration along the $z$ axis, the density controls
the threshold shear force. Although this is inessential, for
simplicity we assume that the ratio of shear to normal stress, i.e.
the friction coefficient, increases linearly with density. As
mentioned earlier, the texture of the medium also contributes
to the friction coefficient. However, since we consider only shear in
a fixed orientation, a single scalar parameter combining density and
texture should suffice. This parameter is
called ``density'' for short and is denoted by $\vr(x,y)$. Thus at any
time the state of the medium is characterized by this field.

We determine the shear band (path in the ($x,y$) plane) by the following
three conditions: {\it a)} it is continuous, {\it b)} it spans the
sample in the $x$ direction without overhangs and {\it c)} the sum of
the density along it is minimal among all possible paths satisfying
{\it a)} and {\it b)}. One can recognize that this is the well known
problem of finding the ground state of a directed polymer in a random
potential \cite{dirpol}.

Relative motion of the particles takes place within the shear band
while the rest of the sample remains still. Small movements can
totally rearrange the local structure \cite{Behr,OuaRou} and thus may
induce large changes in the local density. We simplify this complex
behavior by renewing the density $\vr$ only along the shear band,
by independent random values taken from a fixed distribution.
After this, a new shear band is again located as described above. Thus the
shear process consists of a succession of localized slips occurring at
very small time scales. We note that in characterizing this process, 
in the spirit of a continuum modeling, we ignore potential
stress inhomogeneities in the medium. It is a simplifying assumption
of the model to relate the shear band localization only to the density,
and not to the full solution of the local stress distribution.

In order to be able to simulate the above model we discretized it on a
square lattice either with principle axis parallel to $x$ and $y$ and
considering first and second nearest neighbours, or 
tilted by $45^{\rm o}$ considering only nearest neighbours.
Periodic boundary conditions are imposed in
the $y$ direction. Simulations with site and bond versions 
were also carried out leading essentially to the
same results. We consider here square samples with system size
$N\times N$ with $N$ varying from $32$ to $512$. Initially a density
$\vr_i$ (a random number uniformly distributed between 0 and 1) is
assigned to every bond $i$. We define the instantaneous 
shear band as the
spanning directed path along which $\sum\vr_i$ is minimal (
applying the usual transfer matrix method \cite{dirpol}). Once the shear
band is found the bonds belonging to it are assigned
new values taken from the
same uniform distribution as used initially.  We repeat this process
and monitor different properties of the system \cite{BakSn}.

We define the average density $\langle \vr\rangle$ as the mean value
of the density of the sites {\it not} belonging to the shear band. This
definition, as well as our procedure of choosing the least and
changing it, guarantees that the average density is a monotonically
increasing function of time.

The monotonic behavior and the bounded nature of the average density
($\vr \leq 1$) ensure that it has an asymptotic value. In finite
samples this is equal to 1.  In Fig. \ref{Fig_rho_t} we have plotted
the deviation of the average density from this asymptotic value. At
early times ($t/N\lesssim 2$) the rescaled curves go together
independently of the system size; later non-trivial system size
effects can be observed.  The relaxation to the asymptotic value gets
slower as the system size increases.

\bfg
\centerline{\epsfig{file=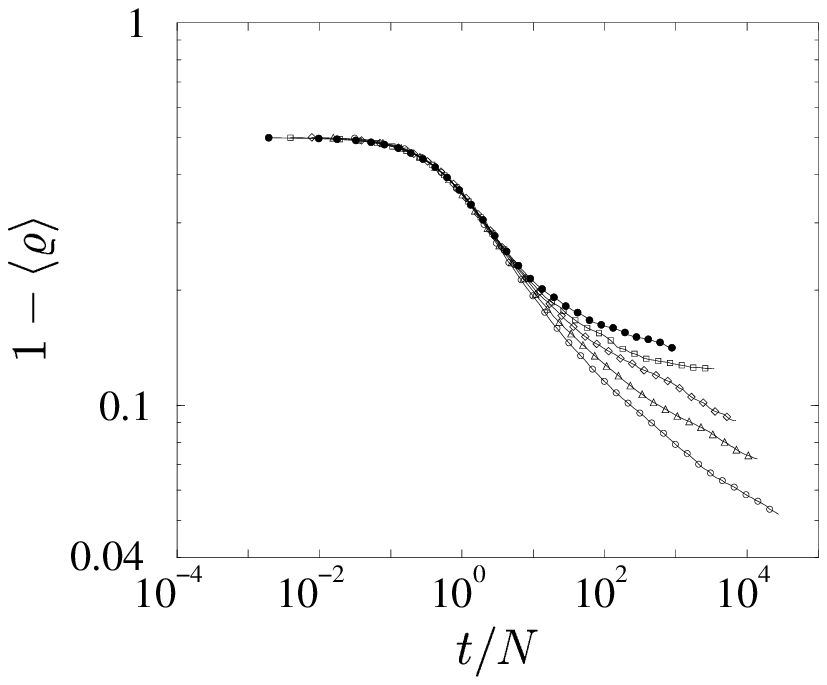,height=6truecm}}
\medskip
\caption{\label{Fig_rho_t}\footnotesize
The difference of the average
density $\langle \vr \rangle$ from its asymptotic 
value is plotted as a function of
time $t$ rescaled by the system size $N$. The system sizes are $32$,
$64$, $128$, $256$ and $512$ from to bottom to top respectively. }
\efg\vskip-10pt

Since the system evolves entirely through the process of
choosing and changing the shear band, 
we have monitored the following two important
quantities related to the shear band: The Hamming distance $d$ (which
is the number of different sites between successive shear bands) (Fig.
\ref{Fig_diff_rhoSB}a) and the average density of the sites along the
shear band $\langle\vr_{SB}\rangle$ before change (Fig.
\ref{Fig_diff_rhoSB}b). It is apparent from the figure that there is a
characteristic time of $t_{c1}\simeq N$, below which the distance is
essentially constant and equal to the system size and the density of
the shear band is roughly constant. This can be understood
qualitatively from the following considerations. Since the very first
shear band is equivalent to the ground state conformation of a
directed polymer in a random potential, we know from this analogy that
the mean density along this shear band is much less than $0.5$
\cite{dirpol}. Once the path is refreshed, its mean density increases
to $0.5$. The next shear band tends to be repelled by the previous one
since there still exist many spanning paths with a lower density.
Thus at early times, two successive shear bands differ completely
(Fig. \ref{Fig_diff_rhoSB}a) and the density of the shear band remains
more or less the same (Fig. \ref{Fig_diff_rhoSB}b). This initial phase
should last until on average all sites have been refreshed a few
times, a number of time steps of the order of $N$.

The absence of overlap between successive shear bands in this early
time regime reflects the fact that no
well defined shear band can be observed in loose granular media. 
Experimentally this is connected to the difficulty in quantifying
fluctuations, when the mean shear strain is of small magnitude.
So what is observed is seemingly a homogeneous shear.

There is a transition regime up to $t_{c2}\simeq 20N$ where we still
have a good quality data collapse. In this regime both curves $d$ and
$r\equiv 0.5-\vr_{SB}$ start to fall off. 
The decreasing distance indicates an increasing persistence of
\bfg
\centerline{\epsfig{file=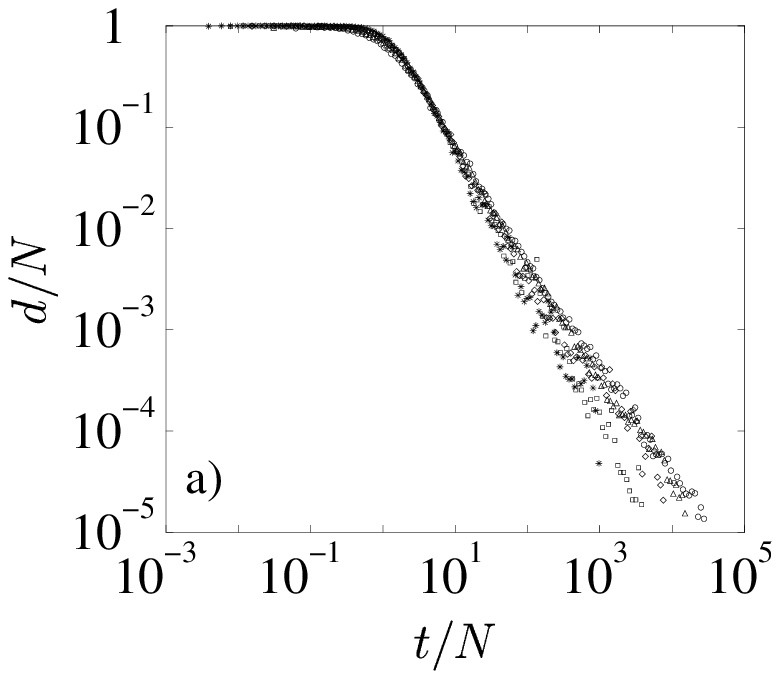,height=4truecm}
\epsfig{file=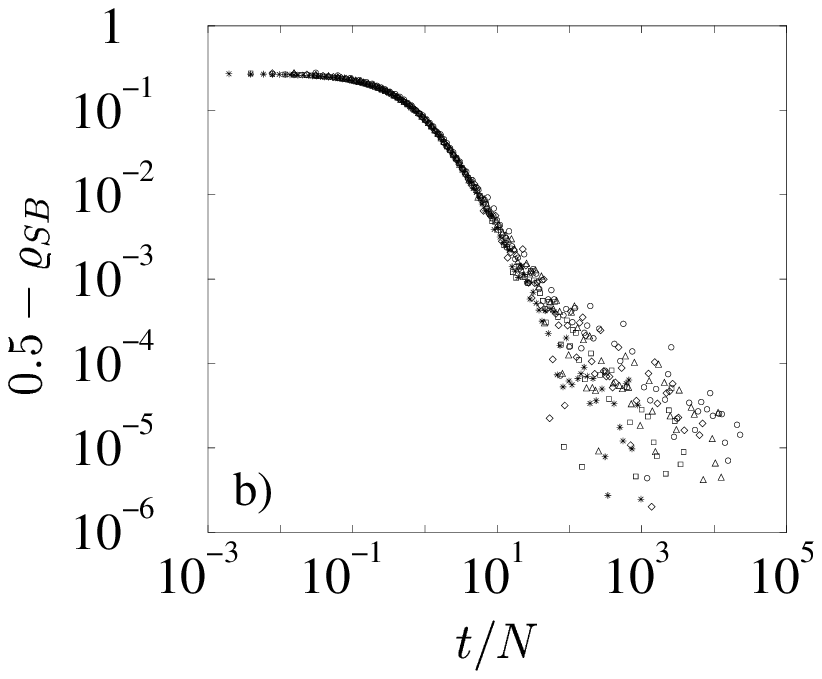,height=4truecm}}
\medskip
\caption{\label{Fig_diff_rhoSB}\footnotesize
a) Log-log plot of the time dependence of the distance $d$ for system
sizes $32$ to $512$ scaled together. Both the distance $d$ and the
time $t$ scale with the system size $N$.  b) Log-log plot of the
deviation of the density of the shear band before the update from its
asymptotic value as a function of time for system sizes $32$ to
$512$ scaled together.
}\efg\noindent\vskip-10pt
the shear band. As the average density of the system increases (Fig.
\ref{Fig_rho_t}) the density of the minimal path also grows and
thus the repulsive interaction between two consecutive shear bands
progressively fades away. Finally, by the end of the transition
regime, the interaction becomes attractive and a much slower
relaxation process takes place.

The above measurements point to a localization of the shear band,
induced by the imposed dynamics. In order to understand better how
this comes about we present density snap shots of the system at
four different instances (Fig. \ref{Fig_maps})
varying from $t/N \sim 4$ to $4000$. We observe that initially
(Fig.  \ref{Fig_maps}a) the density appears
homogeneously distributed. Then progressively high density regions
become apparent. The mechanism for the formation of
these regions is the following: As the average density
increases, the interaction between successive shear bands becomes
attractive and the path gets restricted in space. Small fluctuations
of the shear band then lead to a density increase in this region. The
presence of these surrounding areas of high densities increases the
attraction of successive shear bands, thus leading to a positive
feedback process resulting in regions of finite width and very high
density where the shear band is trapped in the middle, in a
``canyon-like'' structure (black lines surrounded by white in Figs.
\ref{Fig_maps}c and d).

The escape from the above described trap is only possible via a jump
to another local minimum. The probability of such a jump decreases
faster than exponentially with increasing density. Thus as time grows,
the average jump size decreases even though large regions with
relatively small densities remain. The progressive
self-quenching of the shear band in the system is responsible for the
anomalous slow increase in the average density. This inhomogeneous
aging and extremely slow dynamics is reminiscent of a
glassy behavior.

In order to get some more insight into the slow dynamics of the system
we have studied the same model on a hierarchical lattice. The simple
geometry allows for a detailed analytic treatment of the model. This
study will
\bfg
\centerline{\epsfig{file=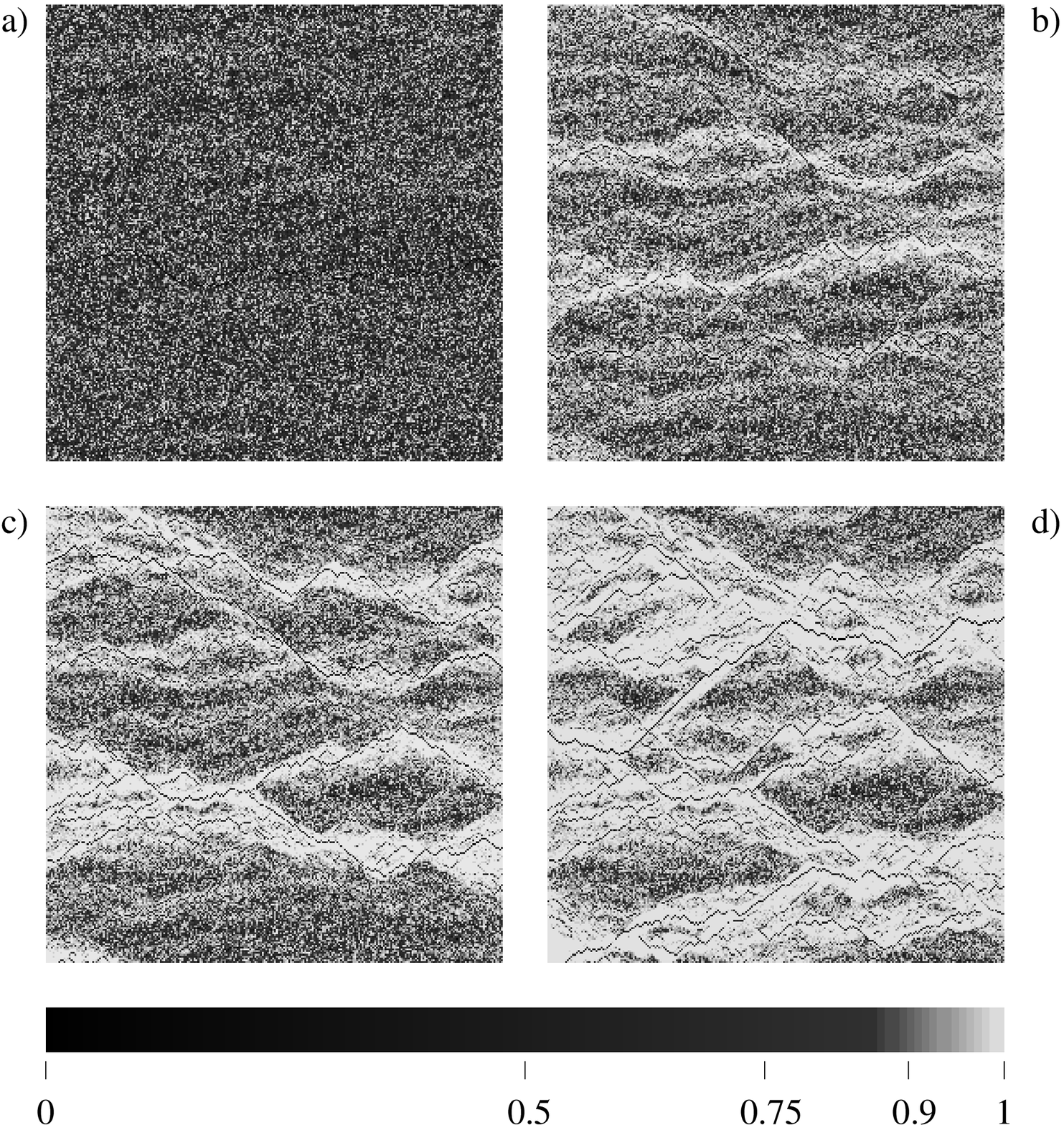,height=9truecm}}
\medskip
\caption{\label{Fig_maps}\footnotesize
Snapshots of densities at different time on a system of size 256 by
256: a) $t=10^{3}\simeq 4N$, b) $t=10^{4}\simeq 40N$, c)
$t=10^{5}\simeq400N$, d) $t=10^{6}\simeq4000N$. The greyscale 
is indicated at the bottom of the figure.
}\efg\noindent\vskip-10pt
be reported elsewhere \cite{TKKR}. Here we only summarize
the main features of this analysis. The slow density increase and
strong system size dependence seen on the square lattice are also seen
in the hierarchical one. Here we can show that $1-\langle\vr\rangle$
decreases as a sum of power-laws with a vanishing exponent depending
on the lattice size, i.e. the number of generations of the
hierarchical lattice. Further, the early time regime is a single
function of $t/N$ as for the square lattice, while the late time
regime scales instead as $t/N^{\alpha}$, where $\alpha = 1/\log(2)$. 

In spite of its simplicity, the model we have introduced displays some
interesting consequences of collective organization of density fluctuations
in a granular assembly. Although only time-independent rules are
introduced, the simulations reveal a slow densification which occurs
together with a non-trivial patterning of the density in the sample.
Simultaneously, the shear strain is localized on shear
bands which acquire progressively a longer and longer persistence. The
occurrence of high density regions confining the shear band is a
feature which should be observable using X-ray tomography
as recently performed in triaxial tests by Desrues {\it et
al}\cite{DCMM}.

{\bf Acknowledgment:} This work was partially supported by OTKA
T024004 and T029985.

\end{multicols}
\end{document}